\documentclass[a4paper,11pt]{article}
\usepackage{jheppub} 

\def\CN{{\cal N}}
\def\CO{{\cal O}}

\def\diag{\mathop{\rm diag}\nolimits}

\def\tr{\mathop{\rm tr}}
\def\rank{\mathop{\rm rank}}

\def\beq#1\eeq{\begin{align}#1\end{align}}

\title{Generalized Hitchin system, Spectral curve and $\mathcal{N}=1$ dynamics}

\author[]{Dan Xie}
\author[]{and Kazuya Yonekura}
\affiliation[]{School of Natural Sciences, Institute for Advanced Study \\
Princeton, NJ 08540, USA}

\abstract{A generalized Hitchin equation was proposed as the BPS equation for a large class of four dimensional $\mathcal{N}=1$ theories engineered using 
M5 branes. In this paper, we show how to write down the spectral curve for the moduli space of generalized Hitchin equations, and extract interesting $\mathcal{N}=1$ dynamics out of it, such as deformed modui space, chiral ring relation, SUSY breaking, etc. Holomorphy plays a crucial role in our construction.}

\begin{document} 
\maketitle
\flushbottom

\section{Introduction}
Seiberg and Witten found the exact solution of  Coulomb branch of $\mathcal{N}=2$ theories by using remarkable physical 
insights~\cite{Seiberg:1994rs,Seiberg:1994aj}. The results can be summarized elegantly using a Seiberg-Witten curve fibered over Coulomb branch. Exact low energy effective action on Coulomb branch is determined by finding a Seiberg-Witten curve. However, the physical methods used in \cite{Seiberg:1994rs,Seiberg:1994aj} are not easy to use for more complicated theories. There are two closely related methods which are used very successfully in finding solutions for general $\mathcal{N}=2$ theories. One is the type IIA brane construction and its M theory lift \cite{Witten:1997sc}, and the other one is using the connection of the Seiberg-Witten solution and integrable system \cite{Donagi:1995cf,Kapustin:1998xn,Gaiotto:2009hg}, in particular, Hitchin system is playing a crucial role in finding solutions. 

It was pointed out by Intriligator and Seiberg~\cite{Intriligator:1994sm} that such curves can also be written down for Coulomb branch of $\mathcal{N}=1$ theories. The curves describe holomorphic gauge couplings of low energy massless $U(1)$ gauge fields. 

Later in the context of the M theory lift of type IIA brane setup, curves are obtained for $\mathcal{N}=1$ Supersymmetric QCD (SQCD)
which are obtained by turning on the mass of the adjoint chiral multiplet inside $\mathcal{N}=2$ vector multiplet \cite{Hori:1997ab,Witten:1997ep}.
A lot of results were obtained  from this approach
(see \cite{Giveon:1998sr} and references therein). 

However, those results are more or less relying on results of $\mathcal{N}=2$ theories which have type IIA brane realization (i.e., linear quiver gauge theories)
and then deforming them by mass of adjoint chiral multiplets, and it is difficult to find the curves for pure $\mathcal{N}=1$ theories, that is,
theories in which adjoint masses are infinity or there are no adjoint fields at all. 
There are many other $\CN=1$ models which have no obvious origin to $\mathcal{N}=2$ linear quiver theories. 
There has been little clue how to write down curves of those models. 

The purpose of this paper is to propose a general method for finding $\mathcal{N}=1$ curves for theories \cite{Xie:2013gma} engineered using M5 branes compactified on a punctured Riemann surface\footnote{ $\mathcal{N}=1$ field theory dynamics of same or similar class are studied in~\cite{Maruyoshi:2009uk, Benini:2009mz, Tachikawa:2011ea, Bah:2011je, Bah:2011vv, Bah:2012dg, Beem:2012yn, Gadde:2013fma,        
Bah:2013qya, Maruyoshi:2013hja, Bah:2013aha, Maruyoshi:2013ega}. 
One of crucial ingredients is the quartic superpotential of \cite{Gadde:2013fma} as we will see.}.
 A generalized Hitchin equation is proposed in \cite{Xie:2013gma} for describing the moduli space of 
 above field theories (see also \cite{Bonelli:2013pva} for another approach to generalized Hitchin systems).
 It is expected that the moduli space of solutions of this generalized Hithcin equations (the moduli space is denoted as  $M_{GH}$) is describing some kind of ``Coulomb''~\footnote{We use quotation mark here because there is no real distinction between the Coulomb and Higgs branches of three dimensional $\mathcal{N}=2$ theory.} branch of the underlying four dimensional theory compactified on a circle. Given the similarity between the generalized Hitchin equations and ordinary Hitchin equations, we expect that a similar spectral curve for $M_{GH}$ could be written which will then describe the holomorphic aspects of $\mathcal{N}=1$ gauge theory dynamics.

Indeed, one can write down a spectral curve for $M_{GH}$, and surprisingingly one can extract  lots of  dynamical information of the low energy dynamics in simply trying to write down the curves. The procedure of determining such curves is surprisingly simple, and the crucial thing is the holomorphy, which agrees with the philosophy taken by Seiberg \cite{Seiberg:1993vc}. 

Let's summarize our main results for determining $\mathcal{N}=1$ curves. There are two Higgs fields $\Phi_1$ and $\Phi_2$  in generalized Hitchin equations. They are sections of $L_1 \otimes {\rm ad}(E)$ and $L_2 \otimes {\rm ad}(E)$ respectively, where $L_1$ and $L_2$ are line bundles 
such that $L_1 \otimes L_2 $ is equal to the canonical bundle $K$, and ${\rm ad}(E)$ is the holomorphic vector bundle  in the adjoint representation of gauge group.
We take fiber coordinates of $L_1$ and $L_2$ as $v$ and $w$, and the coordinates of the Riemann surface where M5 brane wraps is denoted as $z$. These three coordinates parameterize a non-compact local Calabi-Yau manifold~\cite{Bah:2012dg}. 
Our spectral curve is an $N$ cover of the Riemann surface, and it is described by a set of polynomial equations depending on $v,w,z$. 

One can write down two obvious spectral curves for them using the compactification data (punctures and bundles)
\begin{align}
&\det(v-\Phi_1)=0 \rightarrow v^N+\sum_{i=2}^N\phi_{1i}(z)v^{N-i}=0,\nonumber\\
&\det(w-\Phi_2)=0\rightarrow w^N+\sum_{i=2}^N\phi_{2i}(z)w^{N-i}=0.
\end{align}
The coefficients in those two curves are the "Coloumb" branch moduli. If one of the Higgs fields, say $\Phi_2$, vanishes,
the equation for $\Phi_1$ is valid with arbitrary moduli parameters which are consistent with singularities and bundle structures.
It actually gives the moduli space of twisted Higgs bundle and it is shown in \cite{beauville1989spectral,markman1994spectral} that a spectral  curve can be written down. 

When both Higgs fields are nonzero,  there is a crucial commuting condition on those two matrices, 
\begin{align}
[\Phi_1, \Phi_2]=0.
\end{align}
There is a simple fact about the commuting matrices: given a matrix $A$ with generic eigenvalues, 
the matrices commuting with $A$ can be written as a degree $N-1$ polynomial in $A$: $B=h_1 A^{N-1}+h_2 A^{N-2}+\ldots+h_N$, therefore once this link equation is given, the eigenvalues of $B$ are determined by the eigenvalues of $A$. Applying this theorem to our context, 
we need a third equation relating $v$ and $w$ as
\begin{equation}
w=h_{1}(z) v^{N-1}+h_{2}(z) v^{N-2}+\ldots+h_N.
\end{equation}
These three equations are not independent, namely, given the spectral curve of $v$ and this link equation, we should be able to recover the spectral curve of $w$. 
The crucial point is that $h_i$ has to be holomorphic (or meromorphic). The holomorphic property of $h_i$ and the above consistent relation 
put a lot of constraints on various moduli appearing in  spectral equations of $v$ and $w$.

There are many $\mathcal{N}=1$ dynamical informations which can be extracted in the attempt of solving the link equation, and those dynamics do not appear in $\mathcal{N}=2$ context. Let's list some of them: 
\begin{itemize}

\item \textbf{Deformed moduli space}: In some cases, the moduli in spectral curves satisfy the deformed chiral ring relations due to quantum 
effect as in $N_f=N$ SQCD \cite{Seiberg:1994bz}. 
We can recover those type of relations by solving the link equation.
\item \textbf{Chiral ring relation}: In general, the operators in the $v$ and $w$ spectral curves satisfy interesting chiral ring relations 
(including deformed moduli constraints), which can be determined exactly. For example, we find interesting chiral ring relations
for moduli space of Maldacena-Nunez theory \cite{Maldacena:2000mw}.
\item \textbf{Mass deformation}: In $\mathcal{N}=2$ theories, masses of hypermultiplets only change the metric of Coulomb branch, but masses in $\mathcal{N}=1$ theory dramatically change the IR behavior, such as elimination of moduli spaces, SUSY breaking, SUSY restoration, etc.
\item \textbf{Phase structure}: In $\mathcal{N}=2$ case, the curve can be used to probe non-abelian (conformal) and abelian Coulomb phases, and there is always a continuous moduli space. In $\mathcal{N}=1$ case, the curve can
also be used to probe non-abelian and abelian Coulomb phases, and we also find Higgs/Confining phases. In some cases, there are only 
isolated vacua as in pure ${\cal N}=1$ super-Yang-Mills.
\item \textbf{Supersymmetry breaking}: In some cases, one can not find any solution to the link equation, and one can not write a spectral curve. 
Then we conclude that SUSY is dynamically broken~\cite{deBoer:1998by}.\footnote{
However, we cannot immediately say whether there exists a stable SUSY breaking vacuum or the potential is of runaway type.
We need more detailed field theory analysis in this case. 
}
\end{itemize}

This paper is organized as follows: in section~\ref{sec:2}, we discuss how to find the spectral curve of generalized Hitchin system.
In section~\ref{sec:3}, we solve theories engineered using six dimensional $A_1$ theory. 
In section~\ref{sec:SUN}, we solve theories engineered using 6d $A_{N-1}$ theory. Finally, we give a conclusion in section~\ref{sec:concl}.

\section{Generalized Hitchin's equations and spectral curve}\label{sec:2}
\subsection{Generalized Hitchin equation and four dimensional $\mathcal{N}=1$ theory}
Four dimensional $\mathcal{N}=1$ theories can be derived by compactifying six dimensional $(2,0)$ theory
on a punctured Riemann surface. The data defining the theory are
\begin{itemize}
\item A punctured Riemann surface $M_{g,n}$ and a choice of ADE group G.
\item Two line bundles $L_1$ and $ L_2$ such that $L_1\otimes L_2=K$ with $K$ the canonical bundle~\cite{Bah:2012dg}.
The two Higgs fields $\Phi_1, \Phi_2$ are holomorphic sections of $L_1 \otimes {\rm ad}(E)$ and $L_2 \otimes {\rm ad}(E)$ respectively,
where ${\rm ad}(E)$ is the bundle in the adjoint representation of the gauge group. 
\item The local puncture types: a commuting nilpotent pair of G~\cite{Xie:2013gma}.
\end{itemize}
In this paper, we only consider locally $\mathcal{N}=2$ punctures, namely only one of the Higgs fields is singular at a puncture,
with the same types of singularities as in \cite{Gaiotto:2009we,Gaiotto:2009hg}

It is proposed in \cite{Xie:2013gma} 
that the following generalized Hitchin equations are the BPS equations for these $\mathcal{N}=1$
compactifications:
\begin{align}
&D_{\bar{z}} \Phi_1 =D_{\bar{z}} \Phi_2=0, \nonumber\\
&[\Phi_1, \Phi_2]=0, \nonumber\\
&F_{z\bar{z}}+[\Phi_1, \Phi_1^*]h_1+[\Phi_2, \Phi_2^{*}]h_2=0,
\end{align}
Here $h_1$ and $h_2$ are fixed Hermitian metrics for two line bundles $L_1\otimes K^{-1}$ and $L_2\otimes K^{-1}$.

The moduli space of this generalized Hitchin equations is expected to be the target of three dimensional theory derived by compactifying our 
four dimensional theory on a circle, similar to the $\CN=2$ case \cite{Gaiotto:2009hg}. More details about the moduli space will be discussed elsewhere \cite{xie:2013ww}.
The purpose of this paper is to try to use spectral curve to understand this moduli space and therefore 
learn interesting  IR dynamics of field theory. 
\subsubsection{Field theory description and quartic superpotential}
Here let's review the field theory description about those theories constructed from M5 branes. The weakly coupled field theory description 
is described by taking degeneration limit of the Riemann surface. There are two kinds of matter systems; one is represented by a sphere with three regular punctures, and 
the other is described by a sphere with one irregular punctures and a regular puncture. The latter part is useful for describing non-conformal theories.

\begin{center}
\begin{figure}
\small
\centering
\includegraphics[width=12cm]{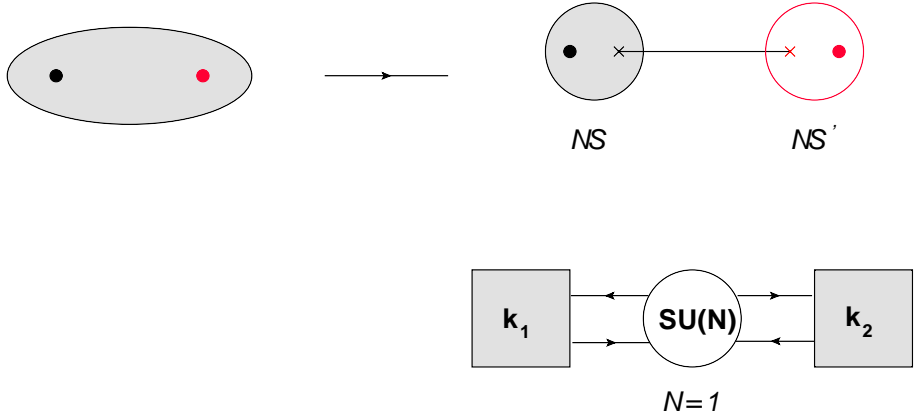}
\caption{A Riemann sphere with two singularities (left), its degeneration limit (upper-right) and the corresponding field theory quiver (lower-right).}
\label{sqcd}
\end{figure}
\end{center}

In the  degeneration limit, there are two types of matter systems which are called NS and NS' matter in \cite{Xie:2013gma}. 
When two matter systems of 
the same type are glued together, we get a $\mathcal{N}=2$ gauge group and a familiar cubic superpotential term involving the moment map and the adjoint chiral field in $\mathcal{N}=2$ vector multiplet. When an NS matter and 
an NS' matter are connected, we get a $\mathcal{N}=1$ gauge group and a quartic superpotential~\cite{Gadde:2013fma}
\begin{equation}
W=c \tr(\mu_1\mu_2),
\end{equation}
here $\mu_1$ and $\mu_2$ are the moment maps for two glued punctures.  Let's give a simple example showing the explicit form of 
the above superpotential term. Our main example in this paper is SQCD which is described by a sphere with one irregular puncture of $\Phi_1$ and 
one irregular puncture of $\Phi_2$. Irregular punctures are
defined to have singularities which are more singular than a simple pole $1/z$ (for more details, see \cite{Nanopoulos:2010zb}).
We only use punctures which can be read off \cite{Gaiotto:2009hg,Gaiotto:2009we} from the solutions of
IIA brane configurations uplifted to M theory \cite{Witten:1997sc}.
They are given as
\begin{align}
& \Phi_1 \to {\zeta\over z^{1+1/(N-k_1)}} \diag (0,\ldots,0, 1,\omega_{N-k_1},\ldots, \omega_{N-k_1}^{N-k_1-1}), ~~(z \to 0), \nonumber\\
& \Phi_2 \to {\zeta z^{1/(N-k_2)}} \diag (0,\ldots,0, 1,\omega_{N-k_2},\ldots, \omega_{N-k_2}^{N-k_2-1}), ~~~ ~~(z \to \infty),
 \end{align}
which describes $k_1$ flavors and $k_2$ flavors separately, where $\omega_{k}=\exp(2\pi i/k)$. 
The irregular punctures are actually the ones used for $\mathcal{N}=2$ SQCD, and we just rotate one of 
the irregular singularity, see figure.~\ref{sqcd}. The bundle structures are $L_1=L_2={\cal O}(-1)$ as we will explain at 
the beginning of section~\ref{sec:3}, and we have taken into account 
$L_2={\cal O}(-1)$ in the above behavior at $z \to \infty$.

The quarks are divided into two sets with $k_1$ and $k_2$ flavors which are represented by a sphere with an irregular singularity and 
a regular singularity. The momental maps for $SU(N)$ gauge groups for these two sets of quarks are
\begin{align}
&(\mu_1)_\alpha^\beta=\sum_{i=1}^{k_1} \left(q_{\alpha}^i \tilde{q}_i^{\beta}-{1\over N} \tr(q_{\gamma}^i \tilde{q}_i^{\gamma})\delta_\alpha^\beta \right),\nonumber\\
& (\mu_2)_\alpha^\beta=\sum_{i=1}^{k_2} \left( p_{\alpha}^i \tilde{p}_i^{\beta}-{1\over N} \tr(p_{\gamma}^i \tilde{p}_i^{\gamma}) \delta_\alpha^\beta \right),
\end{align}
here $\alpha, \beta$ are gauge indices, and $i,j$ are flavor indices. Let's decompose  the meson  as follows:
\begin{equation}
M=\left(
\begin{array}{cc}
M_1  &L\\ 
 \tilde{L} &M_2
\end{array}\right),
\end{equation}
i.e. $M_1$ ($M_2$) represents the meson built from $k_1$ ($k_2$) flavors, and $L, \tilde{L}$ are the mixed mesons constructed by using two sets of quarks.
Then the quartic superpotential is simply
\begin{equation}
W= c\tr(L\tilde{L})-\frac{c}{N} \tr(M_1) \tr(M_2).
\end{equation}

\subsection{Hitchin fibration for moduli space of twisted Higgs bundle}
The Hitchin fibration of the ordinary Hitchin system plays an important role in studying the dynamics of four dimensional $\mathcal{N}=2$ theory. 
In particular, the spectral curve is identified with the Seiberg-Witten curve. 

It is expected that we can also write a spectral curve for the moduli space of generalized Hitchin equations. 
Here we assume $G=SU(N)$ and consider the moduli space of generalized Hitchin  equations on a Riemann surface without punctures (some of the conclusion is also valid when we have punctures). 
There are two special sub-manifolds inside the full moduli space; A: $M_1$ which describes the solutions with $\Phi_1=0~\&~\Phi_2\neq 0$; and
B: $M_2$ which describes the solutions with $\Phi_2=0~\&~\Phi_1\neq 0$. 
These spaces are the so-called moduli space of twisted Higgs bundle \cite{beauville1989spectral,markman1994spectral} , namely the single Higgs field  is not a section of 
canonical bundle but the section of a general bundle $L$.
One can define the Hitchin map on $L_1$ and $L_2$ 
\begin{align}
&\det(v-\Phi_1)=0 \rightarrow v^N+\sum_{i=2}^N\phi_{1i}(z)v^{N-i}=0,\nonumber\\
&\det(w-\Phi_2)=0\rightarrow w^N+\sum_{i=2}^N\phi_{2i}(z)w^{N-i}=0,
\end{align}
where $\phi_{1i}\in H^0(M_{g,n}, L_1^i)$ and $\phi_{2i} \in H^0(M_{g,n}, L_2^i)$.
The Riemann-Roch theorem states 
\begin{equation}
\dim H^0(L)- \dim H^0(L^*\otimes K)=\deg(L)-g+1. 
\end{equation}
Here $\dim H^0(L)$ is the dimension of holomorphic sections of the line bundle $L$.

For example, let's consider $L_1=L_2$, so $\deg(L_1)=\deg(L_2)=g-1$. This is the Maldacena-Nunez theory~\cite{Maldacena:2000mw}.
Then we have $\dim H^0(L^i)=i(g-1)-g+1, i> 2$ and $\dim H^0(L^2)=g$, and the dimension of the Coulomb branch from one spectral curve is 
\begin{equation}
d_b=[\sum_{i=2}^N( i(g-1)-g+1)+1]={1\over2}(g-1)(N^2-N)+1.
\end{equation}
The dimension of the fibre is computed as the genus of the spectral curve if the gauge group is $U(N)$. 
We need to subtract the genus of the base Riemann surface in the $SU(N)$ case. It is given as \cite{beauville1989spectral,markman1994spectral}
\begin{equation}
d_f={1\over 2}(N^2-N)(g-1)+(g-1)(N-1).
\end{equation}
Thus the dimension of the base and the dimension of the fibre are different which is a reflection of $\mathcal{N}=1$ supersymmetry.

\subsection{Commuting matrices}
When both Higgs fields are nonzero, the above two spectral equations are still valid, but there are constraints coming from the 
commuting condition $[\Phi_1, \Phi_2]$, which will relate the specrtra of $\Phi_1$ and $\Phi_2$. Therefore we need a third equation relating $v$ and $w$. The answer is given by the
following simple fact: given two commuting matrices $A$ and $B$, if they
both have distinct eigenvalues, then the matrices $A, B$ could be
written as a degree $N-1$ polynomial of each other 
\begin{align}
&A=f_{N-1} B^{N-1}+f_{N-2}B^{N-2}+\ldots , \nonumber\\
&B=g_{N-1} A^{N-1}+g_{N-2}A^{N-2}+\ldots . \nonumber\\
\end{align}
In particular, the eigenvalues of $A$ are polynomials of eigenvalues of $B$, and vice versa.
In our case, there are families of commuting matrices defined on points of the Riemann surface, and the above formula should be still valid for a generic point
of the Riemann surface. 
The $(v,w)$ in this general case is defined as $N$ pairs of eigenvalues of $(\Phi_1, \Phi_2)$, and hence the above equations are equivalent to
\begin{align}
&v=f_{N-1}(z) w^{N-1}+f_{N-2}(z) w^{N-2}+\ldots , \nonumber\\
&w=g_{N-1}(z) v^{N-1}+g_{N-2}(z) v^{N-2}+\ldots . \nonumber\\
\end{align}
Here $f_i$ and $g_i$ are both meromorphic sections of various line bundles whose degree can be easily found, and we call
them link equations.

Therefore we have three equations: two of them are the spectral curves for $v$ and $w$,
and the third one is the above
link equations relating $v$ and $w$. Given the spectral curve of $v$, we should be able to
write down the spectral curve of $w$ using the above link equation, and vice
verse. Therefore only two equations are independent, and the link equation
is the most important one. This put a lot of constraints on coefficients
of three equations as we will see later explicitly.

In summary, we have the following three polynomial equations which describes the spectral curve of generalized Hitchin system:
\begin{align}
&v^N+\sum_{i=2}^N\phi_{1i}(z)v^{N-i}=0, \nonumber\\
&w^N+\sum_{i=2}^N\phi_{2i}(z)w^{N-i}=0,  \nonumber\\
&w=h_1(z) v^{N-1}+h_2(z) v^{N-2}+\ldots+h_{N-1}(z). \label{eq:linkeq}
\end{align}
The form  of $h_i(z)$ is found by imposing the condition that the combination of the second and third equations imply the first equation. 

\subsubsection{$SU(2)$ case}
In the $SU(2)$ case, the full set of equations can be reduced to the following simple formula
\begin{align}
& v^2=f(z), \nonumber\\
&w^2=g(z), \nonumber\\
& vw=h(z) ,
\end{align}
where $f(z), g(z)$ and $h(z)$ are sections of $L_1^2, L_2^2$ and $K$ respectively. 
In these equations, we have used the fact that $\Phi_1$ and $\Phi_2$ are traceless.
From them, we get $h(z)^2=f(z)g(z)$. This equation puts strong constraint on the coefficients of $f$ and $g$ as 
the square root of their product should be a holomorphic section (or meromorphic at punctures). 
This constraint was also discussed in \cite{Bonelli:2013pva}

\subsubsection{General case: the use of holomorphy}\label{sec:degen}
In the higher rank case, we could derive the coefficients $h_i$ by the requirement that combining the link equation and the spectral 
equation for $v$, we should get the spectral equation for $w$. We would get $N$ nonlinear equations relating $N$ coefficients $h_k$ and $f_k, g_k$, and $h_k$ can be 
uniquely fixed by these equations.
It would be a formidable task to really solve the above nonlinear equations though. However, using the holomorphic (or meromorphic) property of $h(z)$ and the spectral curves of $v$ and $w$,
we can solve $h_k$ and find the constrains among the moduli in spectral equations. Moreover, the link equation can be significantly simplified in some cases. 
In this part, we study the constraints on link equation from the holomorphy of $h_i$.

The most important information about a meromorphic section is its poles and zeros, and we would like to first determine the poles of $h_i$. First of all, 
$h_i$  can not have pole at the pole position of $v$, as otherwise due to the link equation, $w$ would have a pole at the pole position of $v$, which is against our assumption that either $v$ or $w$ has a singularity at a point. Secondly, $h_i$ has to have a pole at the pole position of $w$. 

Thirdly, let us consider an arbitrary point $P$ on Riemann surface where $v$ and $w$ are non-singular here, and  choose a local coordinate such that $P$ has coordinate  $z=0$. Then the spectral curve for $v$ can be expanded as
\begin{align}
&v^N+\sum_{k=2}^{k=N}( \phi_{1,k}(0) +\phi'_{1,k}(0) z+\ldots)v^{N-k}=0  \nonumber\\
\rightarrow~~&\prod_{i=1}^N (v-c_i)+z \sum_{k=2}^{N}a_k v^{N-k}+\ldots=0,
\label{expand}
\end{align}
and the eigenvalues for $v$ near $z=0$ can be expanded as
\begin{equation}
v_i=c_i+z^{\alpha_i}+\ldots.
\label{eigen}
\end{equation}
Here we have defined $\alpha_i=1/n_i$, where $n_i$ is the degeneracy of $c_i$ among $(c_1,\ldots,c_N)$ and we have $\sum n_i=N$. 
We assumed that $\sum_{k=2}^{N}a_k v^{N-k}$ is nonzero at $v=c_i$.
Now let's assume $h_i$ does have a pole at $z=0$. Then near $z=0$, the link equation should have the expansion like, e.g.,
\begin{equation}
w={\prod_{i=1}^{N-1}(v-c_i)\over z}+\ldots,
\label{linkc}
\end{equation}
so as to cancel the pole as far as possible since $w$ is assumed not to have a singularity.  Let's assume that the multiplicities of $(v-c_i)$ in 
above formula is $a_i$, and we have $\sum a_i=N-1$.
Then to cancel the pole completely, we have to impose the following condition
\begin{equation}
{a_i\over n_i}\geq 1.
\end{equation}
However, this is impossible as it implies $\sum a_i\geq \sum n_i$. Therefore $h_i$ is non-singular at the above generic point.  

Finally, we consider a point where $w$ has a singularity but $v$ is regular, and we would like to determine the order of pole of $h_i$. 
Here we do not give a general analysis, and just study a particular important example. Let us assume that the $N$ values of $w$ behave as
\beq
w \sim \left(\frac{1}{z^{1+1/{(N-N_f)}}},\frac{\omega_{N-N_f}}{z^{1+1/{(N-N_f)}}},\ldots, \frac{\omega_{N-N_f}^{N-N_f-1} }{z^{1+1/{(N-N_f)}}}, \frac{m_1}{z},\ldots, \frac{m_{N_f}}{z} \right),
\eeq
where $\omega_{N-N_f}=\exp(2 \pi i/(N-N_f))$.
This is a standard behavior for singularities which have type IIA construction.
To reproduce this singular behavior, we need to have $h_i \sim 1/z^2$ because $w$ is more singular than $1/z$ and $v$ is not singular.
Furthermore, the curves should be given as
\beq
&(v-c_0)^{N-N_f} \sum_{i=1}^{N_f} (v-c_i)+z \sum_{k=2}^{N}a_k v^{N-k}+\ldots=0, \label{eq:degeneracycond}\\
&w=\frac{(v-c_0)^{N-N_f-1} \sum_{i=1}^{N_f} (v-c_i)}{z^2}+\ldots.
\eeq
In this form, the correct behavior of $w$ is reproduced, assuming $\sum_{k=2}^{N}a_k v^{N-k} \neq 0$.
We can use the equation for $v$ to get
\beq
z(v-c_0)w= \sum_{k=2}^N h'_k v^{N-k},\label{eq:simplelink}
\eeq
where $h_k'$ is now non-singular.

Let's first apply the above result to the theory defined on a Riemann surface without any puncture. Without loss of generality, we assume 
$0\leq \deg(L_2)\leq \deg(L_1)$, since if $\deg L_2<0$, we can only have $w=0$.  
Now, under the assumption of genericity $\sum_{k=2}^{N}a_k v^{N-k} \neq 0$ discussed above,
 $h_i$ is a holomorphic section of a line bundle whose degree is $\deg(L_2)-(N-i)\deg(L_1)$, so
 only $h_{N}$ and possibly $h_{N-1}$ have non-negative degree and therefore nonzero.
 Taking into account the traceless condition, the link equation is simplified as
\begin{equation}
w=h_{N-1} v,\label{eq:proportional}
\end{equation}
with $h_{N-1}$ constant for $\deg L_1=\deg L_2$ and $h_{N-1}=0$ for $\deg L_2 <\deg L_1$.
We will also discuss nongeneric case in section \ref{sec:3}.

Next let us consider the case of a Riemann sphere with two punctures at $z=0, \infty$.
In this paper, we are mainly interested in SQCD which is defined on sphere with an irregular singularity for $v$ and $w$ separately, and the bundle structure is $L_1=L_2={\cal O}(-1)$. We take the 
above geometry as our example, and the analysis for more general geometry may be similar. 
We assume that $w$ is singular at $z=0$, so we have \eqref{eq:simplelink}.
We also assume that there is one singularity of $v$ whose singular behavior near $z=\infty$ is $v'= {1\over z^{'1+\gamma}}$ ($z^{'}=1/z, v'=zv$). 
 It is obvious that the $h'_k$
in \eqref{eq:simplelink} should be constant since terms with positive powers of $z$ 
would imply that $w$ is singular at $z=\infty$, which is against our assumption.
Furthermore, by the same reason, only the terms $h_{N-1}'v+h_{N}'$ are allowed.
Therefore, the link equation is given as 
\begin{equation}
w={1\over z} {a v+b\over v-c},\label{eq:linkSQCD}
\end{equation}
where $a=h_{N-1}'$, $b=h_N'$ and $c=c_0$ are constants.

\section{SU(2) theory}\label{sec:3}
The basic formula is the following
\begin{align}
& v^2=f(z), \nonumber\\
&w^2=g(z), \nonumber\\
& vw=h(z) ,
\end{align}
with $h^2=fg$. Typically, we first write down the first two equations using the bundle and puncture structures, and then simply require
that their product is a square of holomorphic section (or meromorphic at punctures). This will link the parameters in $f$ and $g$.

\subsection{SU(2) SQCD}\label{sec:SU(2)SQCD}
We are going to study SU(2) SQCD in detail. There are three types of punctures we are going to use
\begin{align}
& \Phi_A={1\over z^{1+1/2}} \diag (\zeta, -\zeta)+\ldots, \label{eq:N2typeA} \\
& \Phi_B={1\over z^{2}} \diag (\zeta, -\zeta)+{1\over z} \diag(m,-m)\ldots, \label{eq:N2typeB}\\
& \Phi_C={1\over z} \diag(m,-m), \label{eq:N2typeC}
\end{align}
where these expressions are written in the case that the position of the puncture is at $z = 0$.
Here the puncture of type $A$ describes zero fundamental, $B$ describes one fundamental, and $C$ is a regular full puncture. 
The line bundles $L_1$ and $L_2$ of the two scalars $\Phi_1$ and $\Phi_2$ are both 
the bundle $\CO(-1) $ of the sphere, i.e., $\deg L_1=\deg L_2=-1$. 

The fact $\deg L_1=\deg L_2=-1$ can be seen as follows.
Let us consider a brane setup of the SQCD as in \cite{Witten:1997ep,Hori:1997ab}. We prepare two NS5 branes, which we denote as NS5 and NS5', and we suspend $N$ D4 branes between them.
The Higgs field $\Phi_1$ has a pole at NS5 and $\Phi_2$ has a pole at NS5'. 
Originally, $v$ and $w$ are flat coordinates in the brane set up, but as was done by Gaiotto \cite{Gaiotto:2009we},
introducing the poles at the two ends of the D4 branes change the bundle structures.
In the ${\cal N}=2$ case of \cite{Gaiotto:2009we}, 
introducing poles of $v$ at both ends of the D4 branes makes $v$ a coordinate of the canonical bundle $K=\CO(-2)$ of the sphere,
since the coordinate\footnote{In \cite{Gaiotto:2009we}, the notation $x$ is used instead of $v$
for the coordinate of the canonical bundle. Here we continue to use $v$ for nontrivial bundles.} $v$ in the patch near $z=0$ and $v'$
in the patch near $z =\infty$
is related as $v'=z^2v$ in the coordinate change $z \to z'=1/z$.
The $w$ remains to be a coordinate of the trivial bundle.
In our $\CN=1$ case, introducing a pole of $v$ at one end of the D4 branes and a pole of $w$ at 
the other end makes both $v$ and $w$ be coordinates of the $\CO(-1)$ bundle.
Actually, by requiring that $v$ has the appropriate pole at $z=\infty$ and is smooth at $z=0$, we get $\deg L_1 \geq -1$.
In the same way, we get $\det L_2 \geq -1$. Taking into account $\deg L_1+\deg L_2=\deg K=-2$, we obtain $\deg L_1=\deg L_2=-1$.
Roughly speaking, a puncture of $\Phi_1$ (or $\Phi_2$) adds $-1$ to the degree of the line bundle $L_1$ (or $L_2$).

\subsubsection{Pure SU(2)}
There are one $A$ type puncture of $\Phi_1$ at $z=0$ and one $A$ type puncture of $\Phi_2$ at $z=\infty$.  
The two spectral curves without imposing any conditions are \footnote{Notice that we can use scale invariance to make sure $\zeta_1=\zeta_2$, so there is only one independent dimensional parameter in SQCD case. } 
\begin{align}
&v^2={\zeta_1^2 \over z^3}+{u_1\over z^2}, \nonumber\\
& w^2=\zeta_2^2 z+u_2, \nonumber\\
& vw=h(z),
\end{align}
where $h(z)^2=({\zeta_1^2 \over z^3}+{u_1\over z^2} )(\zeta_2^2 z+u_2)$. It is easy to see that in order for $h(z)$ to be meromorphic, 
we need to impose $u_1=u_2=0$, and we get a curve 
\begin{align}
&v^2={\zeta_1^2 \over z^3} \nonumber\\
& w^2=\zeta_2^2 z \nonumber\\
& vw=\zeta_1 \zeta_2/z
\end{align}
which is in agreement with the solution found by Witten \cite{Witten:1997ep} (One need to redefine the coordinates $v$ and $w$).

\subsubsection{SU(2) with one flavor}
There are one $B$ type puncture of $\Phi_1$ and one $A$ type puncture of $\Phi_2$. The curves for $v$ and $w$ are
\begin{align}
&v^2={\zeta_1^2 \over z^4} +{\zeta_1 m\over z^3}+{u_1\over z^2},\nonumber\\
& w^2=\zeta_2^2 z+u_2, \nonumber\\
& vw=h(z).
\end{align}
Again, $h(z)^2=({\zeta_1^2 \over z^4} +{\zeta_1 m\over z^3}+{u_1\over z^2})(\zeta_2^2 z+u_2)$. The holomorphy of $h(z)$ will ensure that 
$u_1=0, u_2=\zeta_1 \zeta_2^2/m$, and our final curve would be
\begin{align}
&v^2={\zeta_1^2 \over z^4} +{\zeta_1 m\over z^3},\nonumber\\
& w^2=\zeta_2^2 z+\zeta_1\zeta_2^2/m, \nonumber\\
& vw={\zeta_2 (\zeta_1{m})^{1/2}\over z^2}(z+{\zeta_1 \over m}).
\end{align}
In the massless limit, $u_2$ is infinity, which means that there is no way to write a meromorphic spectral curve, and we conclude that SUSY is broken. 
This matches perfectly with the field theory fact that the Affleck-Dine-Seiberg superpotential \cite{Affleck:1983mk} is generated and there is no supersymmetric vacuum.

\subsubsection{SU(2) with two flavors}
There are two ways of realizing two flavor theories.  
We can divide the flavor number as $N_f=N'_{f}+N''_f$. Then we have the choice
$N'_f=N''_f=1$ or $N'_f=2,~N''_f=0$.
They are different theories because of the existence of 
quartic superpotential \cite{Gadde:2013fma,Xie:2013gma},
\beq
W=c \tr \left[ 
\sum_{i=1}^{N'_f}\left(q_i \tilde{q}_i-\frac{1}{N}\tr(q_i \tilde{q}_i) \right)   
\sum_{a=1}^{N''_f} \left(p_a \tilde{p}_a-\frac{1}{N}\tr(p_a \tilde{p}_a) \right) \right]. \label{eq:quarticsuper}
\eeq

\paragraph{First realization}:
The puncture types of the first arrangement $N'_f=N''_f=1$ are: there are one $B$ type puncture of $\Phi_1$ and one $B$ type puncture of $\Phi_2$.
The curves are
\begin{align}
&v^2={\zeta_1^2 \over z^4}+{\zeta_1 m_1\over z^3}+{u_1\over z^2}, \nonumber\\
& w^2=\zeta_2^2 z^2 +{\zeta_2 m_2 z}+u_2.
\end{align}
The third equation requires that
\begin{equation}
h(z)^2={1 \over z^4}(u_1 z^2+{\zeta_1 m_1}z +\zeta_1^2)(\zeta_2^2 z^2+{m_2 \zeta_2}z+{u_2 }).
\end{equation}
There are two ways of satisfying this equation. 
The first way is to impose that
\beq
(\zeta_2^2 z^2+{m_2 \zeta_2}z+{u_2 }) \propto (u_1 z^2+{\zeta_1 m_1}z +\zeta_1^2).
\eeq
This condition gives
\begin{equation}
u_1u_2=\zeta_1^2 \zeta_2^2,~~~m_2u_1={\zeta_1 \zeta_2 m_1}. \label{eq:nf=1+1chiral1}
\end{equation}
The second way is to impose
\beq
&(\zeta_2^2 z^2+{m_2 \zeta_2}z+{u_2 }) =(\alpha z+\beta)^2, \\
&(u_1 z^2+{\zeta_1 m_1}z +\zeta_1^2) =(\gamma z+\delta)^2.
\eeq
This condition gives
\beq
 u_2=\frac{m_2^2}{4},~~~u_1=\frac{m_1^2}{4}. \label{eq:nf=1+1chiral2}
\eeq
In either case, there are discrete vacua as expected.

In the massless limit, the condition \eqref{eq:nf=1+1chiral1} becomes
\begin{equation}
u_1u_2=\zeta_1^2 \zeta_2^2. \label{eq:nf=1+1chiral1mless}
\end{equation}
On the other hand, the condition \eqref{eq:nf=1+1chiral2} is 
\beq
u_1=u_2=0. \label{eq:nf=1+1chiral2mless}
\eeq
We may interpret these equations as follows. In the field theory, the low energy superpotential is given by
\beq
W=X(M_{11}M_{22}-M_{12}M_{21}-B\tilde{B}-\Lambda^4)+c (M_{12}M_{21}-\frac{1}{2}M_{11}M_{22}),
\eeq
where $M_{ij}$ are mesons, $B$ and $\tilde{B}$ are baryons, and $X$ is a Lagrange multiplier.
The first term is the deformed moduli constraint of SQCD, 
and the second term comes from the quartic superpotential \eqref{eq:quarticsuper}.
One can see that there are three branches;
\beq
(1):~&M_{11}M_{22}=\Lambda^4, ~~~ M_{12}=M_{21}=B=\tilde{B}=0, ~~~X=\frac{c}{2}, \\
(2):~&M_{12}M_{21}=-\Lambda^4, ~~~ M_{11}=M_{22}=B=\tilde{B}=0, ~~~X=-c, \\
(3):~&B\tilde{B}=-\Lambda^4, ~~~ M_{11}=M_{22}=M_{12}=M_{21}=0, ~~~X=0.
\eeq
The first branch may correspond to \eqref{eq:nf=1+1chiral1mless} by identifying $u_1 \sim (M_{11})^2$ and $u_2 \sim (M_{22})^2$
while the other two branches may corresponds to \eqref{eq:nf=1+1chiral2mless}. See \cite{Yonekura:2013mya} for details about these points.

\paragraph{Second realization}:
In the realization $(N'_f,N''_f)=(2,0)$ of the $N_f=2$ theory, 
there are three punctures: two regular punctures $C$ of $\Phi_1$ at $z=0,\infty$ and a type $A$ puncture of $\Phi_2$ at $z=1$.  
The two curves read
\begin{align}
&v^2={m_1^2\over z^2}+{u_1\over z} +m_2^2,\nonumber\\
& w^2={\zeta^2\over (z-1)^3}+{u_2\over (z-1)^2}.
\end{align}
The third equation requires
\beq
h(z)^2={ 1 \over z^2(z-1)^4} (m_2^2z^2+u_1z+m_1^2)(u_2 (z-1)+\zeta^2)(z-1).
\eeq
This constraint is satisfied if and only if
\beq
u_1=-(m_1^2+m_2^2),~~~(m_2^2-m_1^2)u_2=\zeta^2m_2^2.
\eeq

In the massless limit, $u_1$ has to be zero, and we only get the second curve 
\beq
w^2={\zeta^2\over (z-1)^3}+{u_2\over (z-1)^2},
\eeq
with $u_2$ a free modulus. 
In the field theory side, the flavor symmetry contains a subgroup $SU(2)_1 \times SU(2)_2 \subset SU(4)$, in the spirit of \cite{Gaiotto:2009we}. 
We have chiral operators (usually called mesons)
$\mu_1$ and $\mu_2$ which are in the adjoint representations of $SU(2)_1$ and $SU(2)_2$, respectively.
The deformed moduli constraint is given as $\tr \mu_1^2-\tr \mu_2^2=\Lambda^4$. This is just a rewriting of the usual SQCD 
deformed moduli constraint~\cite{Seiberg:1994bz}.
We may identify the $u_2$ in the curve as the flavor invariant operator of the field theory, $u_2 \sim \tr \mu_1^2+c \sim \mu_2^2+c'$,
where $c$ and $c'$ are constants.
After some coordinate changes, the curve is the same as the ones found in \cite{Intriligator:1994sm,Tachikawa:2011ea}
by taking some of the dynamical scales of \cite{Intriligator:1994sm,Tachikawa:2011ea} to be zero.
There $u_2$ was really identified as $u_2 \sim \tr \mu_1^2+c \sim \mu_2^2+c'$.

\subsubsection{SU(2) with three flavors}
 There are three punctures: two regular punctures $C$ and a type $B$ puncture.
\begin{align}
&v^2={m_1^2\over z^2}+{u_1\over z} +m_2^2,\nonumber\\
& w^2={\zeta^2\over (z-1)^4}+{\zeta m_3\over (z-1)^3}+{u_2\over (z-1)^2}.
\end{align}
The third equation is
\beq
h(z)^2={1 \over z^2(z-1)^4} (m_2^2z^2+u_1z+m_1^2)(u_2 (z-1)^2+m(z-1)+\zeta^2).
\eeq
There are two ways to satisfy this equation.
The first way is
\beq
(m_2^2z^2+u_1z+m_1^2) \propto (u_2 (z-1)^2+m_3(z-1)+\zeta^2).
\eeq
This condition gives
\beq
u_2(u_1+2m_2^2)=m_2^2m_3,~~~\zeta^2(u_1+2m_2^2)=(u_1+m_1^2+m_2^2)m_3.
\eeq
The second way is
\beq
&(m_2^2z^2+u_1z+m_1^2)=(\alpha z+\beta)^2, \\
&(u_2 (z-1)^2+m_3(z-1)+\zeta^2)=(\gamma z +\delta)^2.
\eeq
This condition gives
\beq
u_1^2=\frac{m_1^2 m_2^2}{4},~~~u_2=\frac{4m_3^2}{\zeta^2}.
\eeq

In the massless limit, we must have $u_1=0$. Then the curves are given by $v=0$ and 
\beq
w^2={\zeta^2\over (z-1)^4}+{u_2\over (z-1)^2}.
\eeq
In the field theory side, the low energy superpotential is given as follows.
We take quarks as $q_{i_1i_2\alpha}~(i_1,i_2, \alpha=1,2)$ and $p_\alpha, \tilde{p}_\alpha$.
The $q$ quarks are in the trifundamental representation of $SU(2)_1 \times SU(2)_2 \times SU(2)_g$, where $SU(2)_1 \times SU(2)_2$
are flavor groups and $SU(2)_g$ is the gauge group. We define gauge invariant operators as
\beq
&(\mu_1)_{i_1}^{j_1}=\frac{1}{2}q_{i_1i_2\alpha}q^{j_1i_2\alpha},~~~(\mu_2)_{i_2}^{j_2}=\frac{1}{2}q_{i_1i_2\alpha}q^{i_1j_2\alpha},  \nonumber \\
&N=p_\alpha \tilde{p}^\alpha,~~~,B_{i_1 i_2} =q_{i_1i_2\alpha} p^\alpha,~~~\tilde{B}_{i_1 i_2} =q_{i_1i_2\alpha} \tilde{p}^\alpha,
\eeq
where indices are raised and lowered by the totally antisymmetric tensor.
The quartic superpotential is given as
\beq
W=c ( q_{j_1i_2\alpha}q^{j_1j_2\beta} ) (p_\beta \tilde{p}^\alpha-\frac{1}{2} \delta_\beta^\alpha (\tilde{p}^\gamma p_\gamma))=cB_{i_1 i_2}\tilde{B}^{i_1 i_2}.
\eeq
Including the dynamical superpotential, we obtain
\beq
W=\frac{1}{\Lambda^3} 
\left[N(\tr \mu_1^2 -\tr \mu_2^2)+B_{i_1 i_2}\tilde{B}^{j_1j_2}\left((\mu_1)^{i_1}_{j_1} \delta^{i_2}_{j_2}-\delta^{i_1}_{j_1} (\mu_2)^{i_2}_{j_2} \right)
\right]
+cB_{i_1 i_2}\tilde{B}^{i_1 i_2}.
\eeq
There is a branch such that $B_{i_1i_2}=\tilde{B}_{i_1i_2}=N=0$ and $\tr \mu_1^2=\tr \mu_2^2$.
We may identify $u_2 \sim \tr \mu_1^2=\tr \mu_2^2$.

\subsubsection{SU(2) with four flavors}
The spectral curves are
\begin{align}
&v^2= {m_1^2\over z^2}+{u_1\over z} +m_2^2,\nonumber\\
& w^2={m_3^2\over (z-1)^2}+{u_2 \over (z-1)(z-\tau)}+{m_4^2\over (z-\tau)^2}.
\end{align}
Again, one can fix $u_1$ and $u_2$ using the factorization condition $v^2w^2=h(z)^2$. 
In the massless limit, $u_1=0$ or $u_2=0$.
The curve is at the conformal point if $u_1=u_2=0$. 
One can assign $U(1)_R$ charges as $v:1/2,~w:1/2,~u_1:1,~u_2:1$.
This symmetry is the $U(1)_R$ symmetry of the superconformal algebra.

\subsection{Three irregular singularities: three gauge groups}\label{sec:N2threeiregular}

In this subsection, we study theories which are realized as a Riemann sphere with three punctures.
Suppose that there are $n_1$ punctures of $\Phi_1$ and $n_2$ punctures of $\Phi_2$ with $n_1+n_2=3$.
Then, the degrees of the line bundles $L_1$ and $L_2$ are given by $(\deg L_1,\deg L_2)=(-n_1,1-n_2)$ (or
$(\deg L_1,\deg L_2)=(1-n_1,-n_2)$). 

Let's suppose $(\deg L_1,\deg L_2)=(-n_1,1-n_2)$. This formula may be interpreted as a result of the fact that 
the degree of the line bundle $L_i$ is decreased $-1$ as we introduce a puncture of $\Phi_i$
as discussed at the beginning of section~\ref{sec:SU(2)SQCD}.
The term $1$ in $\deg L_2=1-n_2$ is interpreted as the Euler number of a sphere with three holes around the punctures.
Note that if $(n_1,n_2)=(0,3)$, we get $(\deg L_1,\deg L_2)=(0,-2)$ as it should be for ${\cal N}=2$ theories.

The field theory interpretation of punctures are as follows. The punctures of $\Phi_2$ are the same as in ${\cal N}=2$ theories. 
We have a trifundamental chiral field $Q_{i_1 i_2 i_3}$
of $SU(2)_1 \times SU(2)_2 \times SU(2)_3$ (or more generally a copy of the $T_N$ theory with $SU(N)_1 \times SU(N)_2 \times SU(N)_3$
flavor symmetry), 
and if a puncture is the irregular singularities
of type A, \eqref{eq:N2typeA}, 
the corresponding $SU(2)$ is gauged by ${\cal N}=2$ vector multiplet.
If it is type B, \eqref{eq:N2typeB}, there is an additional fundamental flavor of quarks coupled to the vector multiplet.

The punctures of $\Phi_1$ are kind of ``dual'' of the above ${\cal N}=2$ punctures in the following sense \cite{Gadde:2013fma,Xie:2013gma}.
If the puncture is a regular puncture of type C, \eqref{eq:N2typeC}, then we have a meson $(M_1)^{i_1}_{j_1}$ coupled to the trifundamental field as
\beq
W=\tr(M_1 \mu_1),~~~~~(\mu_1)_{i_1}^{j_1}=Q_{i_1i_2i_3}Q^{j_1 i_2 i_3}, \label{eq:mesoncoupling}
\eeq
where we have assumed that the puncture corresponds to $SU(2)_1$. This is similar to a superpotential in Seiberg dual; 
see \cite{Gadde:2013fma} for details.
If the puncture is the irregular singularity of type A,
the corresponding $SU(2)_1$ is coupled to ${\cal N}=2$ vector multiplet with the superpotential $\tr (M_1 \phi_1)$, where $\phi_1$ is
the adjoint chiral multiplet of the $\CN=2$ vector multiplet. 
By integrating out these massive $M_1$ and $\phi_1$, the end result is that the $SU(2)_1$ is coupled to ${\cal N}=1$
vector multiplet. The puncture of type B can be treated in a similar way. 
It is straightforward to extend the above discussion to the $T_N$ theory.

\subsubsection{Field theory analysis}
We study $SU(2)$ theory with three gauge groups, which is defined by a sphere with three irregular singularities of type A.
Before going into the details of the curve, let us sketch what happens in field theory. We only consider the case $n_1 \geq 1$,
and assume that $SU(2)_3$ is gauged by ${\cal N}=1$ vector multiplet.
Suppose the dynamical scale of the gauge group $SU(2)_3$ is very large. Then,
at low energies, we get a deformed moduli space
\beq
W=X(\tr \mu_1^2-\tr \mu_2^2-\Lambda_1^4),
\eeq
where $X$ is a Lagrange multiplier and $\mu_1$, $\mu_2$ are defined as in \eqref{eq:mesoncoupling}.
We consider the following three cases.

\paragraph{A: $(n_1,n_2)=(3,0),~L_1=\CO(-3),~L_2=\CO(1)$.} 
In this case, we couple ${\cal N}=1$ vector multiplets to $SU(2)_1 \times SU(2)_2$ in addition to the $\CN=1$ gauge group $SU(2)_3$.
The fields $\mu_1$ and $\mu_2$ are in the adjoint representations of the corresponding gauge groups.
Their vevs break $SU(2)_1 \times SU(2)_2 $ to $U(1)^2$. There is one modulus field $u \equiv \tr \mu_1^2=\tr \mu_2^2+\Lambda_1^4$.
Therefore, the low energy theory consists of two massless $U(1)$ fields and one massless modulus $u$.
This is precisely the theory studied in \cite{Tachikawa:2011ea}.
See \cite{Maruyoshi:2013hja} for a generalization to the $T_N$ theory.

\paragraph{B: $(n_1,n_2)=(2,1),~L_1=\CO(-2),~L_2=\CO(0)$.} 
In this case, we couple ${\cal N}=1$ vector multiplet to $SU(2)_1$ and ${\cal N}=2$ vector multiplet to $SU(2)_2$.
The effective superpotential is
\beq
W=X(\tr \mu_1^2-\tr \mu_2^2-\Lambda_1^4)+\tr (\mu_2 \phi_2).
\eeq
Equations of motion of $\phi_2$ set $\mu_2=0$. Then the deformed moduli constraint  requires that 
$\mu_1=\diag(\Lambda_1^2, -\Lambda_2^2)$ up to gauge transformations. This vev breaks $SU(2)_1$ to $U(1)$.
The term $\tr (\mu_2 \phi_2)$ can be regarded as a mass term for $\mu_2$ and $\phi_2$. After integrating out them,
the gauge group $SU(2)_2$ becomes a pure ${\cal N}=1$ Yang-Mills and confines at low energies.
There are no moduli fields. Therefore, the result is that there are two discrete vacua generated by the gaugino condensation of $SU(2)_2$ 
and each vacua has a massless $U(1)$ vector multiplet.

\paragraph{C: $(n_1,n_2)=(1,2),~L_1=\CO(-1),~L_2=\CO(-1)$.} 
In this case, we couple ${\cal N}=2$ vector multiplets to $SU(2)_1 \times SU(2)_2$.
The superpotential is given as
\beq
W=X(\tr \mu_1^2-\tr \mu_2^2-\Lambda_1^4)+\tr (\mu_1 \phi_1)+\tr (\mu_2 \phi_2).
\eeq
Equations of motion of $\phi_1$ and $\phi_2$ set $\mu_1=\mu_2=0$, which is inconsistent with 
the deformed moduli constraint. Thus we conclude that the supersymmetry is broken.
This case is almost the same as the dynamical supersymmetry breaking model of \cite{Izawa:1996pk,Intriligator:1996pu}.
See \cite{Maruyoshi:2013ega} for a generalization to the $T_N$ theory.

\subsubsection{Curves}
Now we would like to discuss the spectral curves of the above theories.

\paragraph{A: $(n_1,n_2)=(3,0),~L_1=\CO(-3),~L_2=\CO(1)$.} 

The curves are
\begin{align}
&v^2={\zeta_1^2\over z^3(z-1)^2}+{\zeta_2^2\over z^2(z-1)^3}+{\zeta_3^2\over z(z-1)^2}+{u\over z^2(z-1)^2} \nonumber\\
&w^2=u_1 z^2+u_2 z+u_3
\end{align}
By looking at the behaviors at $z=0,1,\infty$, one can find that $u_1=u_2=u_3=0$ to satisfy the constraint $v^2w^2=h(z)^2$.
Thus $w=0$ and 
we get a single curve with one moduli $u$. This is in perfect agreement with the Seiberg-Witten curve found in \cite{Tachikawa:2011ea}
after a redefinition $v'=z(z-1)v$.

\paragraph{B: $(n_1,n_2)=(2,1),~L_1=\CO(-2),~L_2=\CO(0)$.} 
The curves are
\begin{align}
&v^2={\zeta_1^2\over z^3}+{u_1\over z^2}+{\zeta_3^2\over z} \nonumber\\
&w^2={\zeta_2^2\over (z-1)^3}+{u_2\over (z-1)^2}+{u_3\over(z-1)}+u_4
\end{align}
Using the link equation $v^2w^2=h(z)^2$, one can fix all the parameters:
\begin{equation}
u_1=-\zeta_1^2-\zeta_3^2,~~u_2=\zeta_2^2+\frac{\zeta_2^2 \zeta_3^2}{\zeta_3^2-\zeta_1^2},
~~u_3=\frac{\zeta_2^2 \zeta_3^2}{\zeta_3^2-\zeta_1^2},~~u_4=0,
\end{equation}
and the curves are
\beq
&v^2=\frac{(\zeta_3^2 z- \zeta_1^2)(z-1)}{z^3}, \nonumber \\
&w^2=\frac{\zeta_2^2}{\zeta_3^2 -\zeta_1^2}\frac{(\zeta_3^2 z- \zeta_1^2)z}{(z-1)^3}, \nonumber \\
&vw=\left(\frac{\zeta_2^2}{\zeta_3^2 -\zeta_1^2} \right)^{1/2} \frac{(\zeta_3^2 z- \zeta_1^2)}{z(z-1)}
\eeq
There are two isolated vacua in this theory corresponding to the square root. This is the same as the prediction from gauge theory.
These curves should give the Seiberg-Witten curve of the massless $U(1)$ field discussed above.
Actually, one can check that the genus of the curve is $1$, consistent with the fact that there is one massless $U(1)$ field at low enegies.

\paragraph{C: $(n_1,n_2)=(1,2),~L_1=\CO(-1),~L_2=\CO(-1)$.} 
The curves are
\beq
&v^2=z\zeta_3^2+ u_1, \nonumber \\
&w^2=\frac{\zeta_1^2}{z^3(z-1)^2}+\frac{\zeta_2^2}{z^2(z-1)^3}+\frac{u_2 z^2+u_3 z+u_4}{z^2(z-1)^2}.
\eeq
By looking at the behavior of $v^2w^2=h(z)^2$ at $z \to 0$ and $z \to 1$, one can check that
there is no solution at all. This is consistent with the supersymmetry breaking in the field theory.

\subsection{Chiral ring relation for Maldacena-Nunez theory }
It is not hard to do calculation for general theory defined on a sphere, and one can also do the calculation on higher genus case by using 
the explicit information about holomorphic sections of various bundles. The general procedure is exactly like what we have done earlier. 
Here we will consider $A_1$ Maldacena-Nunez theory~\cite{Maldacena:2000mw} to show the interesting chiral ring relation one can find by writing down the curve .

Maldacena-Nunez (MN) theory is defined by taking a genus $g$ Riemann surface without any puncture, and the bundle is $L_1= L_2=K^{1\over2}$, where $K$ is the canonical bundle.  We assume that the complex structure of the Riemann surface is such that
it is a hyperelliptic surface for genus $g$ Riemann surface, and the algebraic curve defining it is
\begin{equation}
y^2=\prod_{k=1}^{2g+2}(z-p_k),~~~~p_k\neq p_j.
\end{equation} 
The basis for degree one holomorphic differential (holomorphic sections of the canonical bundle) is
\begin{equation}
e_j={z^j dz \over y},~~~j=0,\ldots,g-1,
\end{equation}
and the basis for holomorphic quadratic differential is 
\begin{align}
&s_j={z^j dz^2 \over y^2},~~~j=0,\ldots, 2g-2, \nonumber\\
&t_j={z^j dz^2 \over y},~~~j=0,\ldots, g-3,
\end{align}
Notice that we have the following simple relation for the product of the degree one differential
\begin{equation}
e_j e_k =e_{j^{'}}e_{k^{'}}=s_{j+k},~~~~j+k=j^{'}+k^{'}.
\end{equation}

Let's first study MN theory on genus two Riemann surface. The three curves are
\begin{align}
& v^2= v_0 e_0+ v_1 e_1, \nonumber\\
& w^2=u_0 e_0+u_1 e_1, \nonumber\\
& v w= h_0 e_0+h_1 e_1.
\end{align}
The consistency condition gives us the following equations
\begin{equation}
h_0^2=v_0 u_0,~~~h_1^2=v_1 u_1,~~~2h_0 h_1 = v_0 u_1+v_1 u_0 .
\end{equation}
The consistency of above equations implies that 
\begin{equation}
v_0 u_1= v_1 u_0.
\end{equation}
This is the interesting chiral-ring relation we found for $A_1$ MN theory.

Let's now generalize the above consideration to $A_1$ MN theory on a genus $g$ Riemann surface, and the curves are
\begin{align}
& v^2= \sum_{i=0}^{g-1} v_i e_i, \nonumber\\
& w^2=\sum_{i=0}^{g-1} u_i e_i, \nonumber\\
& v w= \sum_{i=0}^{g-1} h_i e_i.
\end{align}
The consistency equations are
\begin{align}
& \sum_{i+j=p} h_i h_j= \sum_{i+j=p} v_i u_j,~~~p=0,1,\ldots, 2g-2.
\label{MN}
\end{align}
There are only $g$ constant $h_i$ and there are $2g-1$ equations, so generically we need to impose $(g-1)$ constraints on $v_i$ and $u_i$, which would give us 
the chiral ring relation. The dimension of the moduli space is $g+1$ generically.  One generic branch is given by \eqref{eq:proportional},
$v \propto w$, which leads to
\begin{equation}
v_i u_j= v_j u_i.
\end{equation}
This is the chiral-ring relation relating the parameters in spectral curve of $v$ and $w$.

There are several other branches found by solving equation \eqref{MN}. We can impose chiral ring relation for the parameters inside the spectral curve of $v$ and $w$. 
For example, in the genus three case, we have two branches, one of them is given by
\begin{align}
&v^2= v_0 e_0+ v_1 e_1+v_2 e_2, \nonumber\\
&v^2= u_0 e_0+u_1 e_1+u_2 e_2,  \nonumber\\
& v_0 u_1=v_1 u_0,~~v_1 u_2=v_2 u_1,~~v_2 u_0=v_0 u_2.
\end{align}
and the other one is given by
\begin{align}
&v^2= v_0^2 e_0+ 2v_0v_1 e_1+v_1^2 e_2, \nonumber\\
&v^2= u_0^2 e_0+2u_0u_1 e_1+u_1^2 e_2 . \nonumber\\
\end{align}
Here $v_0, v_1, u_0, u_1$ are arbitrary. 

One can understand those other branches as follows. First take an arbitrary line bundle $L$. Next, take 
two sections of $L$, $s,t \in H^0(L)$, and a section of $K \otimes L^{-2}$, $u \in H^0(K \otimes L^{-2})$.
We can find spectral curves as
\beq
v^2=s^2 u, ~~~
w^2=t^2 u, ~~~
vw=stu. 
\eeq
Therefore there are many branches in the Maldacena-Nunez theory corresponding to the choice of $L$.
The moduli space of this branch is given by 
\beq
\left(H^0(L) \oplus H^0(L) \oplus H^0(K \otimes L^{-2}) \right)/{\mathbb C}^*
\eeq
where division by ${\mathbb C}^*$ means to divide by the equivalence relation 
$(s,t,u) \cong (\alpha s, \alpha t, \alpha^{-2} u)$ for $\alpha \in {\mathbb C}^*$.

\section{ SU(N) theory}\label{sec:SUN}
In this section, we are going to study several interesting examples for $SU(N)$ theories to show how our general procedure of writing down the 
curve can be implemented in practice.

\subsection{SQCD}
\begin{center}
\begin{figure}[htbp]
\small
\centering
\includegraphics[width=14cm]{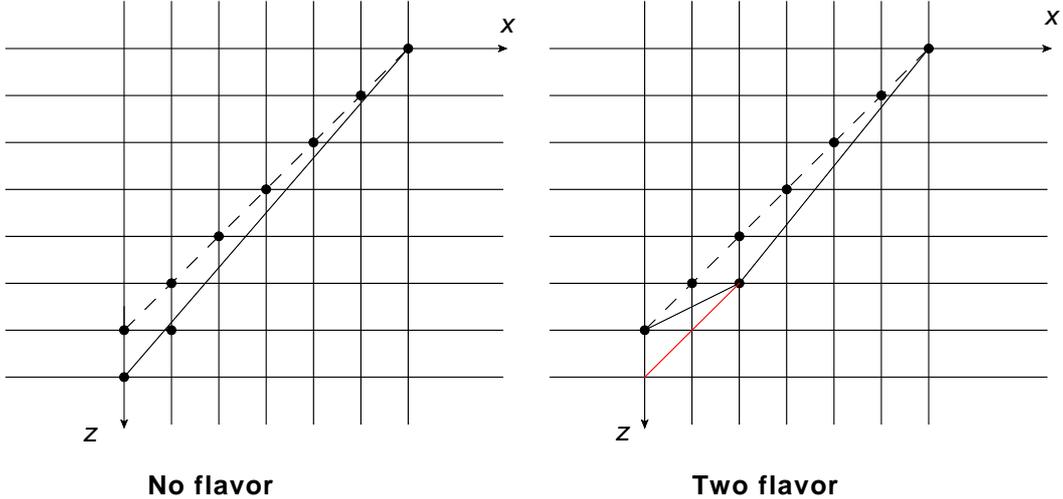}
\caption{The newton polygon for irregular singularity. The red line represents the mass-deformed theory, and the black one represents the massless theory. }
\label{newton}
\end{figure}
\end{center}

In this section, we study a few examples of $SU(N)$ theories. 
The irregular punctures we will use are 
\begin{align}
N_f=0:&~~ \Phi_A={\zeta\over z^{1+1/N}} \diag (1, \omega_N, \omega_N^2,\ldots, \omega_N^{N-1})+\ldots, \label{eq:SUNtypeA} \\
N_f=k:&~~ \Phi_B={\zeta\over z^{1+1/(N-k)}} \diag (0,\ldots, 1,\omega_{N-k},\ldots, \omega_{N-k}^{N-k-1}) \nonumber \\
&~~~~~~~~~~+{1\over z} \diag(m_1,m_2, \ldots, m_k, m,\ldots,m)\ldots,~~ \label{eq:SUNtypeB}\\
N_f=N-1:&~~ \Phi_C={\zeta \over z^{2}} \diag (1,1,1, \ldots, -(N-1)) \nonumber\\
 &~~~~~~~~~~+{1\over z} \diag(m_1,m_2,\ldots,m_{N-1},-(m_1+\ldots+m_{N-1}),) , \label{eq:SUNtypeC}
\end{align}
where $\omega_{k}=\exp(2 \pi i/k)$. The curves for these irregular singularities can be easily read from a Newton polygon, see figure. \ref{newton}. 
The slop of the boundary of Newton polygon encodes the pole orders of the irregular singularity, and the integer lattice points inside the 
Newton polygon represents the deformation in Seiberg-Witten curve. For more details, see \cite{Xie:2012hs}. 

\subsubsection{ $N_f<N_c$ massive SQCD}

Let's consider masssive theory. Let's put one singularity representing $N_f$ flavor at $z=\infty$, and the other singularity 
representing zero flavor at $z=0$. The curves look like
\begin{align}
&v^N+\sum_{i=2}^{N}p_i v^{N-i}+\zeta_1^{N-N_f}z f(v,m)=0, \nonumber\\
& w^N+\sum_{i=2}^{N}{q_i\over z^i} w^{N-i}+ {\zeta_2^N \over z^{N+1}}=0,
\end{align}
where $f(v,m)=\prod_{i=1}^{N_f} (v-m_i)$. 

We use the result of section \ref{sec:degen}.
The condition \eqref{eq:degeneracycond} is satisfied
at $z=0$ only if we set all $p_i=0$. 
By imposing \eqref{eq:degeneracycond} also at $z=\infty$, we have the simple curves
\begin{align}
& v^N+\zeta_1^{N-N_f}z f(v,m_i)=0, \nonumber\\
& (w-u/z)^{N-N_f}g(z,u_i)+{\zeta_2^N \over z^{N+1}}=0,
\end{align}
where $g(z,u_i)=\prod_{i=1}^{N_f} (w-u_i/z)$ and $(N-N_f)u+\sum_{i=1}^{N_f} u_i=0$. 

The link equation \eqref{eq:linkSQCD} can be simplified by simple observation that the two equations look very similar: 
here we take the ansatz that the relation is given by $(w-u/z) v=az^{-1}$. Then after substituting this equation into the second one, we have
\begin{align}
0&=\left( a \over z \right)^{N-N_f}\prod_{i=1}^{N_f} (a/z- v (u_i -u)/z)+{\zeta_2^N \over z^{N+1}} v^N \nonumber \\
\to 0&=v^N+z \left[ \zeta_2^{-N} a^{N-N_f}\prod_{i=1}^{N_f}(u-u_i) \right ] \prod_{i=1}^{N_f} \left(v -\frac{a}{u_i-u} \right).
\end{align} 
Then $u_i$ and $a$ can be found as 
\begin{align}
\frac{a }{u_i-u}=m_i,~~~\left[ \zeta_2^{-N} a^{N-N_f}\prod_{i=1}^{N_f}(u-u_i) \right ] =\zeta_1^{N-N_f}.
\end{align}
The solution is 
\begin{equation}
a^{N} = (-1)^{N_f}\zeta_1^{N-N_f} \zeta_2^{N} \prod_{i=1}^{N_f}m_i  ,~~~u=-\frac{a}{N}\sum_{k=1}^{N_f} \frac{1}{m_k},
~~~u_i-u={a\over m_i}.
\end{equation}
One can easily check that this solution reduces to the solutions of $N_f=0$ \cite{Witten:1997ep} in the limit $m \to \infty$, $\zeta_1 \to 0$ with $a$ fixed.
One can also check that this curve is the same as the one in \cite{Hori:1997ab} which was derived for $m_1=\cdots=m_{N_f}$.

\subsubsection{General massless case}

As we argued in section~\ref{sec:2}, 
the two curves in the massless case before imposing any constraints are
\begin{align}
&(v-v_0)^{N-N_2} \prod_{i=1}^{N_2}(v-v_i)+\zeta_1^{N-N_1}z v^{N_1}=0, \nonumber\\
& (w-w_0/z)^{N-N_1}\prod_{i=1}^{N_1}(w-w_i/z)+{\zeta_2^{N-N_2} \over z^{N-N_2+1}} w^{N_2}=0,
\end{align}
The link equation takes the form
\begin{equation}
w={1\over z}({a v+b \over v -v_0}),
\end{equation}
$a, b$ are fixed by familiar condition that combining link equation and the equation of $v$ would give the equation of $w$. 
A rather singular limit is when $a=b=v_0=0$, then we have to present our link equation in following form:
\begin{equation}
wv=0.
\end{equation}
We will see that both the general and special equations are useful in finding curves for SQCD. 

\paragraph{Generic link equation.}
Let's first consider the generic link equation and
simply substitute the link equation into the equation of $w$. We get 
\begin{equation}
((a-w_0)v+b +v_0 w_0)^{N-N_1}\prod_{i=1}^{N_1}((a-w_i)v+b +w_i v_0)+{\zeta_2^{N-N_2} \over z} (av +b)^{N_2}(v-v_0)^{N-N_2} .
\label{consistent}
\end{equation}
Comparing this equation with the equation of $v$ to make the above equation vanish, we must have the  following conditions:
\begin{align}
& v_i=-b/a, \nonumber\\
&w_i=r,~~b+rv_0=0,~~~a=w_0.
\end{align}
Using the above condition except the $b+rv_0=0$, the curve now looks like
\begin{align}
&(v-v_0)^{N-N_2} (v+b/w_0)^{N_2}+\zeta^{N-N_1}z v^{N_1}=0,\nonumber\\
& (w-w_0/z)^{N-N_1}(w-r/z)^{N_1}+{\zeta^{N-N_2} \over z^{N-N_2+1}} w^{N_2}=0,
\end{align}
with the following condition from the traceless condition of the eigenvalues of $v, w$ :
\begin{equation}
 N_2b/w_0 = (N-N_2) v_0,~~ N_1r=-(N-N_1)w_0. 
 \end{equation}
However, this is consistent with $b+rv_0=0$ only if $N_1=N_2$, assuming $v_0, w_0$ are nonzero. 

Now the parameters $a,b,v_i, w_i=r$ are expressed in terms of $v_0$ and $w_0$.
We get a relation between $v_0$ and $w_0$ by matching equation \eqref{consistent} to equation of $v$, and we have
\begin{equation}
(v_0 w_0)^{N-N_1}=\left(\frac{N_1}{N} \right)^{N} \zeta_1^{N-N_1}\zeta_2^{N-N_2}.\label{eq:v0w0relation}
\end{equation}

The result is consistent with the following field theory branch. As in section~\ref{sec:2},
mesons are given as
\begin{equation}
M=\left(
\begin{array}{cc}
M_1  &L\\ 
 \tilde{L} &M_2
\end{array}\right),
\end{equation}
where $M_{1,2}$ is the $N_{1,2} \times N_{1,2}$ mesons and $L, \tilde{L}$ are $N_{1,2} \times N_{2,1}$ mesons.
Let us consider the branch where $L=\tilde{L}=0$ and $M_{1,2}$ have generic vevs. Then, the effective superpotential for $N \neq N_1+N_2$ is
\beq
W=(N-N_1-N_2) \left ( \frac{\Lambda^{3N-N_1-N_2}}{\det M_1 \det M_2} \right)^{1 \over (N-N_1-N_2)}-\frac{c}{N} \tr M_1 \tr M_2.
\eeq
The first term is the Affleck-Dine-Seiberg superpotential \cite{Affleck:1983mk} for $N_1+N_2<N$. In the case $N_1+N_2>N$,
we consider the Seiberg dual \cite{Seiberg:1994pq} of the theory. Then dual quarks have masses $M_1$ and $M_2$ and can be integrated out
if the rank of the mesons are such that $\rank M=N_1+N_2$.
After integrating out quarks, gaugino condensation generate the above first term. 
The second term $\tr M_1 \tr M_2$ is the quartic superpotential which is present at the tree level.

The equations of motion requires
\beq
&\left ( \frac{\Lambda^{3N-N_1-N_2}}{\det M_1 \det M_2} \right)^{1 \over (N-N_1-N_2)} M_1^{-1}= \frac{c}{N}{\bf 1}_{N_1} \tr M_2, \\
&\left ( \frac{\Lambda^{3N-N_1-N_2}}{\det M_1 \det M_2} \right)^{1 \over (N-N_1-N_2)} M_2^{-1}= \frac{c}{N}{\bf 1}_{N_2} \tr M_1 .
\eeq
By multiplying these equations by $M_1$ or $M_2$ and taking traces, one can see that these equations are consistent
only if $N_1=N_2$. This result is consistent with the above result from the spectral curve. Then, we get 
\beq
M_1=m_1 {\bf 1}_{N_1},~~~~M_2=m_2 {\bf 1}_{N_2},
\eeq
with a constraint
\beq
(m_1m_2)^{N-N_1}=\left( \frac{N}{cN_1N_2} \right)^{N-2N_1} \Lambda^{3N-2N_1}.
\eeq
This precisely matches with \eqref{eq:v0w0relation} by identifying $m_1 \propto v_0$ and $m_2 \propto w_0$.
Although we have assumed $N_1+N_2 \neq N$, a similar analysis gives us the same result for $N_1+N_2 = N$.

\paragraph{Special link equation.}

Next let's  consider  the special link equation $wv=0$. It is easy to see that this is only possible for  $N_1+N_2\geq N$,  because $w$ ($v$) has at least $N-N_1$ ($N-N_2$) nonzero eigenvalues at the punctures, and the sum of the numbers of the non-zero eigenvalues should not exceed $N$. The curves consistent with the link equations are
\begin{align}
&v^{N-k_1}\left(\prod_{i=1}^{k_1}(v-v_i)+\zeta_1^{N-N_1}z v^{N_1+k_1-N} \right)=0, \nonumber\\
& w^{N-k_2} \left( \prod_{i=1}^{k_2}(w-w_i/z)+{\zeta_2^{N-N_2} \over z^{N-N_2+1}} v^{N_2+k_2-N}  \right)=0, \nonumber\\
& wv=0,
\end{align}
where $k_1+k_2=N$, and $\sum_i v_i=\sum_i w_i=0$.

This result is consistent with the field theory. Let us neglect the quartic superpotential for the moment.
Then, for the $N_1+N_2 \geq N$ case, there is a branch of the moduli space in the field theory such that 
$\rank M \leq N$ \cite{Seiberg:1994pq}. This is a classical constraint for $M^i_j=\tilde{q}^i q_j$ since the quarks have at most
rank $N$, and this branch survives quantum mechanically for SQCD with $N_1+N_2 \geq N$.
The maximal possible rank is achieved when $\rank M_1=k_1$ and $\rank M_2=k_2$ with $k_1+k_2=N$.
The inclusion of the quartic superpotential $\tr M_1 \tr M_2$ makes the trace part of these mesons massive
and they can be integrated out. Then we get $\tr M_1=\tr M_2=0$. This condition should correspond to $\sum_i v_i=\sum_i w_i=0$
in the curve. Therefore, $v_i$ and $w_i$ are interpreted as eigenvalues of $M_1$ and $M_2$, respectively.

\subsection{A sphere with three irregular singularities}

As discussed in section~\ref{sec:2}, it is easy to write down the curve if one of the Higgs fields is zero. Then the generalized Hitchin system becomes a twisted Higgs bundle.

Let us consider a simple example: $\Phi_1$ has three irregular singularities of type A at $z=0,1,\infty$
and $L_1=\CO(-3), L_2=\CO(1)$ as in one of the examples of section \ref{sec:N2threeiregular}. We set $\Phi_2=0$. Then there is no constraint on $\Phi_1$, and the curve is completely determined by 
\beq
v^N+\sum_{i=2}^N \frac{u_i v^{N-i} }{z^i(z-1)^i} +\frac{\zeta_1^N}{z^{N+1}(z-1)^N}+\frac{\zeta_2^N}{z^{N}(z-1)^{N+1}}+\frac{\zeta_3^N}{z^{N-1}(z-1)^N}=0.
\eeq
This curve is precisely the same as the Seiberg-Witten curve in \cite{Maruyoshi:2013hja} after a redefinition $v'=z(z-1)v$.

\subsection {Maldacena-Nunez theory}
Let's now consider higher rank Maldacena-Nunez theory, and we will find interesting chiral-ring relation. The bundles are $L_1=L_2=K^{1/2}$ and there are no punctures. 
We have shown in section \ref{sec:2} that the link equation describing one branch can be written down as \eqref{eq:proportional},
\begin{equation}
w=h v,
\end{equation}
here $h$ is just a constant since it is a section of trivial bundle. The spectral curves are written as
\begin{align}
&v^N+\sum f_k v^{N-k}=0, \nonumber\\
&w^N+\sum g_k w^{N-k}=0, \nonumber\\
&w=h v .
\end{align}
Here $f_k$ and $g_k$ can be expanded using a basis $e_i^{(k)}$ of holomorphic sections of $K^{k/2}$ whose dimension is $d_k=(k-1)(g-1)+\delta_{k,2}$:
\begin{align}
& f_k=\sum_{i=1}^{d_k}v_i^{(k)}e_i^{(k)}, \nonumber\\
& g_k=\sum_{i=1}^{d_k}u_i^{(k)}e_i^{(k)} .
\end{align}
Using the consistency condition of the three equations, we can easily find the chiral ring relation
\begin{equation}
v_i^{(k)}u_j^{(k)}=v_j^{(k)}u_i^{(k)}.
\end{equation}

Another branch may be found by taking the ansartz, e.g.,
\begin{align}
& v^N=f(z),~~w^N=g(z),~~~wv=h(z), \nonumber\\
& fg=h^N ,
\end{align}
so we need to choose the parameters inside $f$ and $g$ to make $h$ holomorphic. This will give us lots of interesting chiral-ring relation. 

Finally, we can have the link equation
\begin{equation}
wv=0.
\end{equation}
In this case, the spectral curves of $v$ and $w$ factorizes as 
\begin{align}
& v^{k_1}(v^{N-k_1}+\ldots)=0,
& w^{k_2}(w^{N-k_2}+\ldots)=0,
\end{align}
where $k_1+k_2=N$ and the moduli fields inside the bracket is arbitrary other than the traceless condition.

\section{Conclusion}\label{sec:concl}

We developed a general method for finding $\mathcal{N}=1$ curve for various kinds of theories engineered using M5 branes. 
We find many dynamical properties of the gauge theories in attempting to write down the curves. It is quite remarkable that we simply
start with a mathematically defined spectral curve without using any physical input, and we can recover the highly non-trivial quantum dynamics, such as deformed moduli space,
Seiberg-Witten curve, etc. The perfect agreement with the field theory results clearly shows that our method is correct.

Our basic construction follows from the generalized integrability property of the moduli space of twisted Higgs bundle as shown in \cite{beauville1989spectral,markman1994spectral}. 
It is interesting
that there are also connections to integrable system even for $\mathcal{N}=1$ theory. This rather surprising connection to integrable system 
definitely deserves further study.

It is interesting that we can probe all kinds of phase structures using the spectral curve, and let's summarize the characteristic feature of 
the curves in different phases:
\begin{itemize}
\item \textbf{Non-abelian Coulomb phase}:  This phase typically happens at the origin, i.e. when all the moduli are set to zero. If 
we set all the moduli to be zero, and all the eigenvalues of $v$ and $w$ to be zero, then we can define a $U(1)$ symmetry which 
does not change the fibre and it is identified as the $U(1)_R$ symmetry. Typical examples are $SU(N)$ SQCD with $N_f=2N$ 
theory and Maldacena-Nunez theory.
\item \textbf{Abelian Coulomb phase}: Such phases happens when the genus of the spectral curve 
is non-zero. Typically there are other moduli, but it is possible that the vacua are discrete. In the field theory, there are massless abelian gauge fields. Typical examples are the theory defined using a sphere with two irregular 
punctures and one regular puncture, as first studied by Intriligator-Seiberg \cite{Intriligator:1994sm}.
\item \textbf{Higgs phase}: This phase happens when there is a continuous moduli and the genus of the spectral curve is zero. This happens for many SQCD examples. The quartic superpotential is crucially important to match the spectral curve and the deformed moduli space of the field theory.
\item \textbf{Mass gapped phase}: If there are no moduli in the curve and the genus of the curve is zero, there is no massless fields at low energies.
There are discrete vacua in most cases.
A typical example is the ${\cal N}=1$ pure super-Yang-Mills.
\end{itemize}
We found these phases using the spectral curve. It would be interesting to understand these phases using traditional order parameter, such
as Wilson loop, 't Hooft loop, surface operators \cite{Gaiotto:2013sma} etc. Those extended objects may also have a geometric interpretation in terms of the generalized 
Hitchin equation and M5 brane construction.

In this paper, we only discuss the $\mathcal{N}=1$ dynamics relating to the attempt of writing down the curves. It is definitely interesting 
to extract more physical information out of the curves, such as the position of the singularity where monopoles become massless,
etc. 

We have only studied some particular examples in this paper to show how our construction works nicely. It is interesting to 
work out more examples such as new theories defined using three punctured sphere, and more general theory including $\mathcal{N}=1$ 
Argyres-Douglas theories.  The spectral curve construction presented in this paper is a very important tool, and we believe  that our methods provide 
an extremely powerful way of studying gauge dynamics of four dimensional $\mathcal{N}=1$ theory.

\section*{Acknowledgments}
We would like to thank K. Maruyoshi, Steve Rayan, Y. Tachikawa, Peng Zhao for helpful discussions.
This research is supported in part by Zurich Financial services membership and by the U.S. Department of Energy, grant DE-SC0009988  (DX). 
The work of K.Y. is supported in part by NSF grant PHY- 0969448.

\bibliographystyle{JHEP}
\bibliography{ref}

\end{document}